\documentclass[12pt,preprint]{aastex62}
\usepackage{natbib}
\usepackage{xspace}

\newcommand{\be}{\begin{equation}}
\newcommand{\ee}{\end{equation}}

\newcommand{\Fermi}{\textit{Fermi}\xspace}
\newcommand{\Fermilat}{\textit{Fermi}-LAT\xspace}

\shorttitle{4FGL DR2}
\shortauthors{\Fermilat collaboration}

\begin{document}

 

\title{\Fermi Large Area Telescope Fourth Source Catalog Data Release 2}
    

\author{J.~Ballet}
\email{jean.ballet@cea.fr}
\affiliation{AIM, CEA, CNRS, Universit\'e Paris-Saclay, Universit\'e Paris Diderot, Sorbonne Paris Cit\'e, F-91191 Gif-sur-Yvette, France}
\author{T.~H.~Burnett}
\email{tburnett@u.washington.edu}
\affiliation{Department of Physics, University of Washington, Seattle, WA 98195-1560, USA}
\author{S.~W.~Digel}
\email{digel@stanford.edu}
\affiliation{W. W. Hansen Experimental Physics Laboratory, Kavli Institute for Particle Astrophysics and Cosmology, Department of Physics and SLAC National Accelerator Laboratory, Stanford University, Stanford, CA 94305, USA}
\author{B.~Lott}
\email{lott@cenbg.in2p3.fr}
\affiliation{Centre d'\'Etudes Nucl\'eaires de Bordeaux Gradignan, IN2P3/CNRS, Universit\'e Bordeaux 1, BP120, F-33175 Gradignan Cedex, France}
\collaboration{The \Fermilat collaboration}

\begin{abstract}
We present an incremental version (4FGL-DR2, for Data Release 2) of the fourth \Fermilat catalog of $\gamma$-ray sources.
Based on the first ten years of science data in the energy range from 50~MeV to 1~TeV, it uses the same analysis methods as the 4FGL catalog did for eight years of data. The spectral parameters, spectral energy distributions and associations are updated for all sources. Light curves are rebuilt for all sources with 1-year intervals (not 2-month intervals).

Among the 5064 4FGL sources, 120 are formally below the detection threshold over 10 years (but are kept in the list), while 53 are newly associated and four associations were withdrawn. We report 723 new sources, mostly just above the detection threshold, among which two are considered identified and 341 have a plausible counterpart at other wavelengths.
\end{abstract}


\section{Introduction}
\label{introduction}

The \Fermi Large Area Telescope (LAT) has been surveying the high-energy $\gamma$-ray sky since 2008 \citep{LAT09_instrument} and the LAT Collaboration has published a succession of source catalogs based on comprehensive analyses of LAT data.
The fourth source catalog \citep[4FGL,][]{LAT20_4FGL} was derived from analysis of the first 8 years of LAT science data. Every FGL catalog until and including 4FGL had used a new analysis method, new calibrations, a new diffuse model, and even a new reconstruction of the events themselves. Therefore each successive version of the catalog started from a clean slate and did not use in the processing any information from the previous versions; results for different versions were compared only after the fact.

Since the development of the 4FGL the data have been stable, no new model for the Galactic interstellar emission was built and the analysis method has not evolved. So we decided to change our approach and start building incremental 4FGL versions every two years, until one major analysis or data component changes. This first incremental Data Release (DR2) covers 10 years of data.
This note focuses on what has changed and the way we integrate new sources into 4FGL, and describes the results. The reader is referred to the 4FGL paper for the detailed methodology.

This work updates all the information present in the 4FGL catalog, except the 2-month light curves. Those are very costly (in terms of CPU and disk space) and we have shown in the 4FGL paper that the 1-year light curves capture most of the variability information.

Section~\ref{lat_and_background} describes the data and the updates to the diffuse model, Section~\ref{catalog_main} the updates to the analysis,
Section~\ref{dr2_description} the results, and Section~\ref{dr2_assocs} the updates to the associations.

\section{Instrument \& Background}
\label{lat_and_background}

\subsection{The LAT Data}
\label{LATData}

The data for the 4FGL DR2 catalog were taken during the period 2008 August 4 (15:43 UTC) to 2018 August 2 (19:13 UTC) covering ten years.
During most of this time, \Fermi was operated in sky-scanning survey mode, with the viewing direction rocking north and south of the zenith on alternate orbits such that the entire sky is observed every $\sim$3 hours.
Starting on 2018 March 16 the \Fermi spacecraft was put in safe hold after one solar array drive became stuck. Scientific operations of the LAT were interrupted for more than three weeks, by far the longest missing time interval since 2008, and restarted on April 8 in partial sky-scanning mode\footnote{See \url{https://fermi.gsfc.nasa.gov/ssc/observations/types/post_anomaly/}.}. The Sun no longer enters the field of view, and during some phases of the $\sim$53-day precession period of the orbit, the entire sky is not covered every three hours. This hampers solar observations and short-term light curves, but has little impact on the integrated sky coverage.

As in 3FGL, intervals around solar flares and bright GRBs were excised. During the additional two years, 108 ks were cut due to two successive bright solar flares in 2017 September, and 2.4 ks around 5 new bright GRBs.
The current version of the LAT data remains Pass 8 P8R3 \citep{LAT13_P8, LAT18_P305}.
The energy range remains 50~MeV to 1~TeV, and the data is split over the same 15 components with the same zenith angle selections as in 4FGL.

\subsection{Model for the Diffuse Gamma-Ray Background}
\label{DiffuseModel}

We used exactly the same model for the interstellar emission and the same isotropic spectrum as in 4FGL.
The model for the emission of the Sun and the Moon was also kept the same, neglecting the modulation of the emission along the solar cycle. It became necessary, however, to generate a different effective model for each year (for the light curves). Because of the solar array drive anomaly during the tenth year, that year's coverage of the Sun differs markedly from the previous years.

\section{Construction of the Catalog}
\label{catalog_main}

Most of the steps were identical to 4FGL. The 75 extended source templates were not changed, and no new extended source was added.
We use the same Test Statistic TS = 2 $\log (\mathcal{L} / \mathcal{L}_0)$ to quantify how significantly a source emerges from the background, comparing the maximum value of the likelihood function $\mathcal{L}$ including the source in the model with $\mathcal{L}_0$, the value without the source.
The spectral shapes used for fitting over the entire band were the same, and the source fluxes over seven bands and 1-year light curves were generated in exactly the same way.

\subsection{Detection and Localization}
\label{catalog_detection}

The source detection followed the same lines as in 4FGL.
It used $pointlike$ and a specific diffuse model in which the non-template features are estimated differently.
It started from the 4FGL sources, relocalized them over 10 years of data, looked for peaks in the residual TS maps generated for several spectral shapes, introduced those in the model, refit and iterated over the full procedure.
It resulted in 8946 seed sources at TS $>$ 10.

That procedure naturally resulted in changing the positions of all sources.
Since we wanted the catalog to be incremental, we forced all 4FGL sources back to their original positions (consistent with their names). We applied the same procedure as in 4FGL to eliminate seeds too close to a bright source and inside extended sources, so that 3621 new seeds were entered to the $gtlike$ source characterization in addition to the 5065 4FGL entries.

We reassessed the systematic corrections to localization. Up to 4FGL they were defined globally over the entire sky, so their values were dominated by high-latitude sources (92\% of associated point sources in 4FGL are at $|b| > 5\arcdeg$).
However, confusion and strong interstellar emission make localization much more difficult in the Galactic plane. In 4FGL-DR2 we estimated localization systematics separately at $|b| < 5\arcdeg$, based on 110 pulsars and low-latitude AGN. We found they should indeed be increased there, to $27\arcsec$ for the absolute 95\% error, and 1.37 for the systematic factor. We did not change the systematics in the high-latitude sky ($25\arcsec$ and 1.06).

\subsection{Thresholding, Spectral Shapes and Light Curves}
\label{catalog_significance}

We used the same Science Tools version (v11r7p0) as in 4FGL. The Fermi Tools environment 
was not yet fully stable when the bulk of the DR2 computation was carried out (Fall 2019).
The likelihood weights were recomputed over 10 years of data, resulting in slightly smaller weights throughout.

Adding new seeds forbid using the same RoIs as in 4FGL, so we reoptimized all RoIs, resulting in 1519 RoIs containing up to 10 sources in their core. The maximum of 8 sources in the RoI core enforced in 4FGL was not optimal (it led to too many small neighboring RoIs and too many iterations).

All 4FGL sources were entered in the analysis with their 4FGL spectral model, except three newly identified pulsars and the AGN CTA 102 (4FGL J2232.6+1143) whose main spectral model was changed to \texttt{PLSuperExpCutoff}. CTA 102 is the source with highest TS after the six brightest pulsars and 3C 454.3, another AGN already fit with \texttt{PLSuperExpCutoff}, and changing the spectral model improved the modeling of its surroundings at low energy.

A major difference with the 4FGL procedure is that we protected all 4FGL sources, so that they would not be deleted from the model even if they have TS $<$ 25.
The iteration process was fast, since the sky did not change much with only 25\% more data. The resulting catalog contains 5788 entries, among which 723 are new, and 120 are 4FGL sources at TS $<$ 25. 

Thanks to the improved statistics, more sources are considered significantly curved in 4FGL-DR2 than in 4FGL. The number of \texttt{LogParabola} spectral shapes increased from 1302 to 1572 (among which 126 are new sources), while the number of \texttt{PLSuperExpCutoff} increased from 222 to 230 (all among 4FGL sources). The fraction of sources with a curved spectral model increased from 30 to 31\% overall.

Similarly, the number of significantly variable sources increased from 1443 to 1525 (among which 70 are new sources). The fraction of variable sources decreased because part of the variability information in 4FGL came from the 2-month light curves. The distribution of fractional variability is similar to 4FGL, peaking between 50 and 90\%. The blazars all have fractional variability larger than 10\% except 4C +55.17 (4FGL J0957.6+5523), whose flux decreased by about 13\% over 10 years, not showing any flare.

\subsection{Analysis Flags}
\label{catalog_analysis_flags}

\begin{deluxetable*}{crrrl}

\tablecaption{Comparison of the number of flagged sources between 4FGL and 4FGL-DR2, separately for the 4FGL sources and the new sources. Flag 13 is new to DR2.
\label{tab:flags}}
\tablehead{
\colhead{Flag\tablenotemark{a}} & \colhead{4FGL} & \colhead{4FGL in DR2} & \colhead{New in DR2} & \colhead{Meaning}
}

\startdata
  1  & 215 & 172 & 69 & $TS < 25$ with other model or analysis \\
  2  & 215 & 91 &  5 & Moved beyond 95\% error ellipse\tablenotemark{b} \\
  3  & 342 & 401 & 89 & Flux changed with other model or analysis \\
  4  & 212 & 246 & 104 & Source/background ratio $<$ 10\% \\
  5  & 398 & 414 & 108 & Confused \\
  6  &  92 & 165 & 38 & Interstellar gas clump (c sources) \\
  9  & 136 &  92 & 37 & Localization flag from {\it pointlike} \\
 10  &  27 &  39 & 2 & Bad spectral fit quality \\
 12  & 103 &  81 & 30 & Highly curved spectrum \\
 13  & \nodata & 120 & \nodata & $TS < 25$ at 10 years \\
 All & 1163 & 1252 & 273 & Any flag (Flags $>$ 0) \\
\enddata
 
\tablenotetext{a}{In the FITS file the values are encoded as individual bits in the \texttt{Flags} column, with Flag $n$ having value $2^{(n-1)}$.}
\tablenotetext{b}{
  4FGL had more sources with Flag 2 because it
  relied on the comparison to FL8Y (different interstellar model) but 4FGL-DR2 relies on the comparison to $pointlike$ (\S~\ref{catalog_detection}).}

\end{deluxetable*}

The flags are recalled in Table~\ref{tab:flags} (see the 4FGL paper for the detailed definitions), together with the numbers of sources flagged for each reason and their evolution since 4FGL.
The effect of the underlying interstellar emission model (IEM) was estimated by launching the procedure described in \S~\ref{catalog_significance} a second time using the same seeds but the previous IEM (gll\_iem\_v06).

We note four changes with respect to the 4FGL procedure:
\begin{itemize}
\item For 4FGL, Flag 1 was applied to the comparison with both $pointlike$ and the alternative IEM. For 4FGL-DR2, the comparison to $pointlike$ was performed with no TS selection in order to apply Flag 3 to sources at TS $<$ 25. Flag 1 was determined only by the comparison with the old IEM, after selection at TS $>$ 25 on both sides. 
\item Flag 2 mostly relies on the comparison between the new $pointlike$ positions of the 4FGL sources (\S~\ref{catalog_detection}) and the 4FGL positions, because no localization was performed over 10 years with the old IEM. The comparison with the old IEM also flags a few sources (although the seed positions are exactly the same) when among two nearby seeds one survives in 4FGL-DR2 and the other one with the old IEM.
\item The visual screening for diffuse features (Flag 6) depends on the source parameters, so it has been entirely redone. To be conservative, we have chosen to leave that flag (and the \texttt{c} suffix) in all the sources that had it in 4FGL, even if they would not have been flagged in DR2, and add the flag to all the new cases, including the sources already in 4FGL.
\item We introduce a new Flag (13) to flag explicitly sources with TS $<$ 25 in DR2.
\end{itemize}
Overall the fraction of sources with any flag set increased from 23\% in 4FGL to 26\% in 4FGL-DR2. Among the new DR2 sources, 73\% are flagged in the Galactic plane ($|b| < 10\arcdeg$), but only 20\% above the Galactic plane.

\section{The 4FGL Catalog}
\label{dr2_description}

The catalog is available online\footnote{See \url{https://fermi.gsfc.nasa.gov/ssc/data/access/lat/10yr_catalog/}.}, together with associated products.
It contains 5788 entries (723 new).
The source designation is \texttt{4FGL JHHMM.m+DDMM}.
A new column (\texttt{DataRelease}) is set to 1 for the 4FGL sources and to 2 for the new sources.
Apart from that (and the absence of the 2-month light curves) the format is the same as 4FGL.

\begin{figure}[!ht]
   \centering
   \begin{tabular}{cc}
   \includegraphics[width=0.5\textwidth]{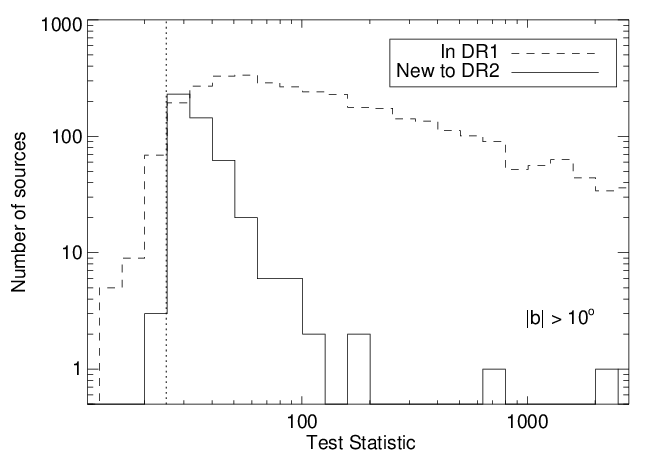} & 
   \includegraphics[width=0.5\textwidth]{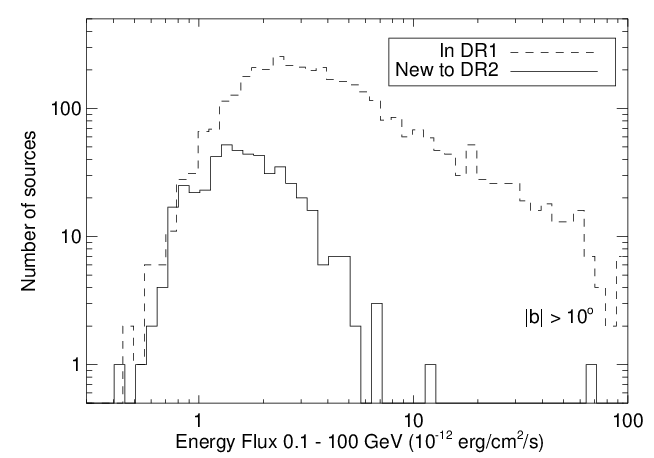}
   \end{tabular}
\caption{Left: Distributions of the 4FGL-DR2 Test Statistic at high-latitude ($|b|>10\degr$) for the sources already in 4FGL (dashed) and those new to DR2 (solid). The vertical dotted line is the threshold at TS = 25. Right: Same distributions for the energy flux.}
\label{fig:TS_enflux}
\end{figure}

\subsection{The 4FGL sources}
\label{4fglsources}

The TS of the 5064 DR1 sources increased on average compared to
4FGL, by 16\% among sources which are fit with the \texttt{PowerLaw model} in both catalogs.
This is less than expected from the exposure increase (24\%). The deficit is not due to the lower weights (it does not depend on index) but to the selection bias (sources tend to be brighter in the interval in which they were defined) and signal-splitting with new sources. The statistical uncertainties on the parameters decreased by 9\%.

The random character of adding new data and the variable nature of the $\gamma$-ray sources inevitably lead to broadening the TS distribution, in addition to shifting it to larger values. The sharp selection threshold at TS = 25 in 4FGL becomes a tail of 120 sources at TS $<$ 25 in 4FGL-DR2 (Figure~\ref{fig:TS_enflux}, left), among which 13 have TS $<$ 16.
The scatter on the TS ratio between DR2 and DR1 is larger in the Galactic plane than at high latitude (real variability is lower in the Galactic plane, but confusion is much larger). One confused Galactic plane source (4FGL J1209.9$-$5923) has TS $<$ 9.

The energy flux decreased on average by 0.15 $\sigma$ between DR1 and DR2. This is again a selection bias dominated by faint sources: close to the detection threshold, DR1 sources can get either brighter or fainter in DR2, but DR2 sources cannot get fainter in DR1 because they would have TS $<$ 25. Real variability is apparent for bright sources (TS $>$ 1000). The scatter on the energy flux ratio is 4.3 $\sigma$ on variable DR1 sources, but only 0.9 $\sigma$ on non-variable ones. There is less scatter on the power-law index (only 0.5 $\sigma$) because source fluxes vary strongly, but spectral shapes only slightly.

\subsection{The new DR2 sources}
\label{dr2sources}

The 723 new DR2 sources are on average very close to the detection threshold (median TS of 32). This is obvious in Figure~\ref{fig:TS_enflux} (left). There are a few newly active bright blazars at TS $>$ 100, but the majority are faint sources that rise above the detection threshold thanks to the additional two years.
Figure~\ref{fig:TS_enflux} (right) shows that the new sources typically have energy fluxes between 1 and $3 \times 10^{-12}$ erg cm$^{-2}$ s$^{-1}$.
Only 70 of the new DR2 sources are considered variable, 64 of them outside the Galactic plane ($|b| > 10\arcdeg$).

\section{Associations}
\label{dr2_assocs}

About 200 counterpart names of DR1 sources have been changed. It was noted that  blazar names from very large surveys (like 2MASS or WISE) were used for some 4FGL associations while more appropriate names from radio catalogs were available.  Moreover some names referred to sources that are  offset by up to a few arcminutes from the real counterpart. We have replaced the non-radio names with those of radio counterparts whenever possible. Note that the positions reported in the (\texttt{RA\_Counterpart}, \texttt{DEC\_Counterpart}) fields were correct.        

Changes in associations of DR1 sources are:
\begin{itemize}
\item Three FSRQs (PKS 0736$-$770, TXS 1530$-$131, PKS 1936$-$623) have been reclassified as BCUs.
\item Recent follow-up observations of 4FGL blazars \citep{Fup_Mas12, Fup_Pag14, Fup_Pen17,Fup_Pen19} have enabled the classification of 132 former BCUs into 118 BL Lacs and 14 FSRQs.
\item The pulsar PSR J1909$-$3744 was mistakenly associated with 4FGL J1912.2$-$3636. The 4FGL source is now unassociated.
\item The SPP sources associated with 4FGL J0129.0+6312 and 4FGL J1102.0$-$6054 have been reclassified as BCUs.
\item The increase of localization systematic uncertainties for sources lying at $ |b|<5 \arcdeg$ (\S~\ref{catalog_detection}) has led to the association of five extra BCUs, five SPPs, and the suppression of the association of 4FGL J1804.9$-$3001  with the globular cluster NGC 6528. Four sources classified as unknown in 4FGL (``unk'', i.e., $|b|<10\arcdeg$ sources associated solely via the Likelihood-Ratio (LR) method from large radio and X-ray surveys) are now classified as blazar candidates of unknown type because they gained Bayesian-based associations. Two other formerly unknown sources are now associated to a pulsar and a binary star (Kleinmann's star).
\item Pulsations have been discovered in six sources, 4FGL J0652.9+4707 (PSR J0653+4706), 4FGL J1221.4$-$0634 (PSR J1221$-$0633), 4FGL J1400.6$-$1432 (PSR J1400$-$1431), 4FGL J1921.4+0136 (PSR J1921+1929), 4FGL J2039.4$-$3616 (PSR J2039$-$3616), and 4FGL J2039.5$-$5617 (PSR J2039$-$5617).   
\item  A total of 24 associations with millisecond pulsars from the West Virginia University list\footnote{\url{http://astro.phys.wvu.edu/GalacticMSPs/GalacticMSPs.txt}} have been added.
\item  The improved position accuracy of the ATNF pulsar PSR J1306$-$4035  has enabled its association with 4FGL J1306.8$-$4035.
\item The latest version of the Radio Fundamental Catalog\footnote{rfc\_2020a available at \url{http://astrogeo.org/rfc/}} has enabled the association of six previously unassociated sources with  blazar candidates.
\item The association of the MSP binary recently discovered by \citet{Assoc20_Wang} with  4FGL J0935.3+0901 has been implemented.
\item Following \citet{Assoc15_Bog}, 4FGL J1544.5$-$1126 is a candidate transitional MSP binary (1RXS J154439.4$-$112820). Its class, ``blazar candidate of unknown type'' in 4FGL, has been changed to ``low-mass X-ray binary''.
\item Another transitional MSP binary, 1SXPS J042749.2$-$670434, displaying simultaneous optical, X-ray, and gamma-ray eclipses has been identified by \citet{Assoc20_Ken} in 4FGL J0427.8$-$6704 (previously unassociated).
\item Six periodic optical and X-ray sources were recently discovered positionally consistent with 4FGL sources. These redback millisecond pulsar candidates have been classified as ``binary''. The 4FGL sources are 4FGL J0212.1+5321 \citep{J0212_Li16}, 4FGL J0523.3$-$2527\citep{J0523_Strader14}, 4FGL J0744.0$-$2525 \citep{J0744_Salvetti17}, 4FGL J0838.7$-$2827 \citep{J0838_Halpern17}, 4FGL J0955.3$-$3949 \citep{J0955_LiHou18}, and 4FGL J2333.1$-$5527 \citep{J2333_Swihart20}.
\item The tentative association of 4FGL J0647.7$-$4418 with the HMB  RX J0648.0$-$4418 called out in DR1 has been replaced by the association with the BCU SUMSS J064744$-$441946 following the multiwavelength investigation of  \citet{Assoc20_Marti}.
\item The association of the pulsar PSR J1757$-$2421 with 4FGL J1756.6$-$2352 has been removed. The pulsar is now associated with the new source 4FGL J1757.9$-$2419.
\end{itemize}

Concerning new DR2 sources, the association procedure  has been performed  by means of the Bayesian and LR methods along the lines of DR1. Associations have been obtained for 342 sources, representing an association fraction of 47\%. This fraction is roughly consistent with that expected for the low-TS sources making up the DR2 sample, based on DR1. 

These associations comprise:
\begin{itemize}
\item one identified pulsar, PSR J1757$-$2421 (see above), and two pulsar candidates, PSR J1439$-$5501, and PSR J1904$-$11;
\item one globular cluster, NGC 362;
\item one high-mass binary, PSR B1259$-$63. Although detected by the LAT during its 2010 periastron passage \citep{LAT11_PSRB1259}, this source is included in an FGL catalog for the first time thanks to the gain in significance from its 2017 passage \citep{Assoc18_Joh};
\item three supernova remnants: 3C 397, SNR G001.4$-$00.1, and SNR G003.7$-$00.2. They are considered as solid associations (as opposed to SPPs, see below) because their radio sizes are quite small (4-14') and their positions well match those of the 4FGL sources;
\item two star-forming regions, NGC 346, which is the brightest star-forming region in  the SMC, and Sh 2-152;
\item two galaxies,  IC 678 and NGC 5380;
\item one starburst galaxy, Arp 299;
\item two radio galaxies, NGC 3078 and NGC 4261; 
\item 283 blazars, including 185 BCUs, 59 BL Lacs, and 39 FSRQs;
\item 15 SPPs (sources of unknown nature but overlapping with known supernova remnants or pulsar-wind nebulae and thus candidates to these classes);
\item 30 sources of unknown nature.
\end{itemize}
We provide low-probability ($0.1<P<0.8$) associations for 36 sources and associations with Planck counterparts for 14 others. 

Only about 40  unassociated sources lie within  $1\fdg5$ of the Galactic plane. Most of the others form broad shoulders (extending beyond 6$\arcdeg$) around the plane, confirming a feature already found in DR1.     

Note: A typo has been found in \S~6.2 of the 4FGL paper concerning the name of the star cluster associated with the extended H\,{\sc ii} region encompassing  4FGL J1115.1$-$6118. The star cluster is NGC 3603 (and not NGC 4603).

\acknowledgments
The \textit{Fermi} LAT Collaboration acknowledges generous ongoing support
from a number of agencies and institutes that have supported both the
development and the operation of the LAT as well as scientific data analysis.
These include the National Aeronautics and Space Administration and the
Department of Energy in the United States, the Commissariat \`a l'Energie Atomique
and the Centre National de la Recherche Scientifique / Institut National de Physique
Nucl\'eaire et de Physique des Particules in France, the Agenzia Spaziale Italiana
and the Istituto Nazionale di Fisica Nucleare in Italy, the Ministry of Education,
Culture, Sports, Science and Technology (MEXT), High Energy Accelerator Research
Organization (KEK) and Japan Aerospace Exploration Agency (JAXA) in Japan, and
the K.~A.~Wallenberg Foundation, the Swedish Research Council and the
Swedish National Space Board in Sweden.
 
Additional support for science analysis during the operations phase is gratefully
acknowledged from the Istituto Nazionale di Astrofisica in Italy and the Centre
National d'\'Etudes Spatiales in France. This work performed in part under DOE
Contract DE-AC02-76SF00515.

This work made extensive use of the ATNF pulsar  catalog\footnote{\url{http://www.atnf.csiro.au/research/pulsar/psrcat}}  \citep{ATNFcatalog}.  This research has made use of the NASA/IPAC Extragalactic Database (NED) which is operated by the Jet Propulsion Laboratory, California Institute of Technology, under contract with the National Aeronautics and Space Administration, and of archival data, software and online services provided by the ASI Science Data Center (ASDC) operated by the Italian Space Agency.
We used the Manitoba SNR catalog \citep{Ferrand2012_SNRCat} to check recently published extended sources.

\software{Gardian \citep{Diffuse2}, HEALPix\footnote{\url{http://healpix.jpl.nasa.gov/}} \citep{Gorski2005}, Aladin\footnote{http://aladin.u-strasbg.fr/}}

\facility{\Fermilat}

\bibliography{Bibtex_4FGL_v1}

\begin{thebibliography}{}
\expandafter\ifx\csname natexlab\endcsname\relax\def\natexlab#1{#1}\fi
\providecommand{\url}[1]{\href{#1}{#1}}
\providecommand{\dodoi}[1]{doi:~\href{http://doi.org/#1}{\nolinkurl{#1}}}
\providecommand{\doeprint}[1]{\href{http://ascl.net/#1}{\nolinkurl{http://ascl.net/#1}}}
\providecommand{\doarXiv}[1]{\href{https://arxiv.org/abs/#1}{\nolinkurl{https://arxiv.org/abs/#1}}}

\bibitem[{{Abdo} {et~al.}(2011){Abdo}, {Ackermann}, {Ajello},
  {et~al.}}]{LAT11_PSRB1259}
{Abdo}, A.~A., {Ackermann}, M., {Ajello}, M., {et~al.} 2011, \apjl, 736, L11,
  \dodoi{10.1088/2041-8205/736/1/L11}

\bibitem[{{Abdollahi} {et~al.}(2020){Abdollahi}, {Acero}, {Ackermann},
  {Ajello}, {Atwood}, {et~al.}}]{LAT20_4FGL}
{Abdollahi}, S., {Acero}, F., {Ackermann}, M., {et~al.} 2020, \apjs, 247, 33,
  \dodoi{10.3847/1538-4365/ab6bcb}

\bibitem[{{Ackermann} {et~al.}(2012){Ackermann}, {Ajello}, {Atwood},
  {et~al.}}]{Diffuse2}
{Ackermann}, M., {Ajello}, M., {Atwood}, W.~B., {et~al.} 2012, \apj, 750, 3,
  \dodoi{10.1088/0004-637X/750/1/3}

\bibitem[{{Atwood} {et~al.}(2009){Atwood}, {Abdo}, {Ackermann},
  {et~al.}}]{LAT09_instrument}
{Atwood}, W.~B., {Abdo}, A.~A., {Ackermann}, M., {et~al.} 2009, \apj, 697,
  1071, \dodoi{10.1088/0004-637X/697/2/1071}

\bibitem[{{Atwood} {et~al.}(2013){Atwood}, {Albert}, {Baldini},
  {et~al.}}]{LAT13_P8}
{Atwood}, W.~B., {Albert}, A., {Baldini}, L., {et~al.} 2013, Fermi Symposium
  proceedings - eConf C121028.
\newblock \doarXiv{1303.3514}

\bibitem[{{Bogdanov} \& {Halpern}(2015)}]{Assoc15_Bog}
{Bogdanov}, S., \& {Halpern}, J.~P. 2015, \apjl, 803, L27,
  \dodoi{10.1088/2041-8205/803/2/L27}

\bibitem[{{Bruel} {et~al.}(2018){Bruel}, {Burnett}, {Digel}, {Johannesson},
  {Omodei}, \& {Wood}}]{LAT18_P305}
{Bruel}, P., {Burnett}, T.~H., {Digel}, S.~W., {et~al.} 2018, $8^{\rm th}$
  Fermi Symposium.
\newblock \doarXiv{1810.11394}

\bibitem[{{Ferrand} \& {Safi-Harb}(2012)}]{Ferrand2012_SNRCat}
{Ferrand}, G., \& {Safi-Harb}, S. 2012, Advances in Space Research, 49, 1313,
  \dodoi{10.1016/j.asr.2012.02.004}

\bibitem[{{G{\'o}rski} {et~al.}(2005){G{\'o}rski}, {Hivon}, {Banday},
  {Wandelt}, {Hansen}, {Reinecke}, \& {Bartelmann}}]{Gorski2005}
{G{\'o}rski}, K.~M., {Hivon}, E., {Banday}, A.~J., {et~al.} 2005, \apj, 622,
  759, \dodoi{10.1086/427976}

\bibitem[{{Halpern} {et~al.}(2017){Halpern}, {Strader}, \&
  {Li}}]{J0838_Halpern17}
{Halpern}, J.~P., {Strader}, J., \& {Li}, M. 2017, \apj, 844, 150,
  \dodoi{10.3847/1538-4357/aa7cff}

\bibitem[{{Johnson} {et~al.}(2018){Johnson}, {Wood}, {Kerr}, {Corbet},
  {Cheung}, {Ray}, \& {Omodei}}]{Assoc18_Joh}
{Johnson}, T.~J., {Wood}, K.~S., {Kerr}, M., {et~al.} 2018, \apj, 863, 27,
  \dodoi{10.3847/1538-4357/aad185}

\bibitem[{{Kennedy} {et~al.}(2020){Kennedy}, {Breton}, {Clark}, {Dhillon},
  {Kerr}, {Buckley}, {Potter}, {S{\'a}nchez}, {Stringer}, \&
  {Marsh}}]{Assoc20_Ken}
{Kennedy}, M.~R., {Breton}, R.~P., {Clark}, C.~J., {et~al.} 2020, \mnras,
  \dodoi{10.1093/mnras/staa912}

\bibitem[{{Li} {et~al.}(2016){Li}, {Kong}, {Hou}, {Mao}, {Strader}, {Chomiuk},
  \& {Tremou}}]{J0212_Li16}
{Li}, K.-L., {Kong}, A. K.~H., {Hou}, X., {et~al.} 2016, \apj, 833, 143,
  \dodoi{10.3847/1538-4357/833/2/143}

\bibitem[{{Li} {et~al.}(2018){Li}, {Hou}, {Strader}, {Takata}, {Kong},
  {Chomiuk}, {Swihart}, {Hui}, \& {Cheng}}]{J0955_LiHou18}
{Li}, K.-L., {Hou}, X., {Strader}, J., {et~al.} 2018, \apj, 863, 194,
  \dodoi{10.3847/1538-4357/aad243}

\bibitem[{{Manchester} {et~al.}(2005){Manchester}, {Hobbs}, {Teoh}, \&
  {Hobbs}}]{ATNFcatalog}
{Manchester}, R.~N., {Hobbs}, G.~B., {Teoh}, A., \& {Hobbs}, M. 2005, \aj, 129,
  1993, \dodoi{10.1086/428488}

\bibitem[{{Mart{\'\i}} {et~al.}(2020){Mart{\'\i}}, {S{\'a}nchez-Ayaso},
  {Luque-Escamilla}, {Paredes}, {Bosch-Ramon}, \& {Corbet}}]{Assoc20_Marti}
{Mart{\'\i}}, J., {S{\'a}nchez-Ayaso}, E., {Luque-Escamilla}, P.~L., {et~al.}
  2020, \mnras, 492, 4291, \dodoi{10.1093/mnras/staa072}

\bibitem[{{Massaro} {et~al.}(2012){Massaro}, {D'Abrusco}, {Tosti}, {Ajello},
  {Paggi}, \& {Gasparrini}}]{Fup_Mas12}
{Massaro}, F., {D'Abrusco}, R., {Tosti}, G., {et~al.} 2012, \apj, 752, 61,
  \dodoi{10.1088/0004-637X/752/1/61}

\bibitem[{{Paggi} {et~al.}(2014){Paggi}, {Milisavljevic}, {Masetti},
  {Jim{\'e}nez-Bail{\'o}n}, {Chavushyan}, {D'Abrusco}, {Massaro}, {Giroletti},
  {Smith}, {Margutti}, {Tosti}, {Mart{\'\i}nez-Galarza}, {Ot{\'\i}-Floranes},
  {Landoni}, {Grindlay}, \& {Funk}}]{Fup_Pag14}
{Paggi}, A., {Milisavljevic}, D., {Masetti}, N., {et~al.} 2014, \aj, 147, 112,
  \dodoi{10.1088/0004-6256/147/5/112}

\bibitem[{{Pe{\~n}a-Herazo} {et~al.}(2017){Pe{\~n}a-Herazo}, {Marchesini},
  {{\'A}lvarez Crespo}, {Ricci}, {Massaro}, {Chavushyan}, {Landoni}, {Strader},
  {Chomiuk}, {Cheung}, {Masetti}, {Jim{\'e}nez-Bail{\'o}n}, {D'Abrusco},
  {Paggi}, {Milisavljevic}, {La Franca}, {Smith}, \& {Tosti}}]{Fup_Pen17}
{Pe{\~n}a-Herazo}, H.~A., {Marchesini}, E.~J., {{\'A}lvarez Crespo}, N.,
  {et~al.} 2017, \apss, 362, 228, \dodoi{10.1007/s10509-017-3208-7}

\bibitem[{{Pe{\~n}a-Herazo} {et~al.}(2019){Pe{\~n}a-Herazo}, {Massaro},
  {Chavushyan}, {Marchesini}, {Paggi}, {Landoni}, {Masetti}, {Ricci},
  {D'Abrusco}, {Milisavljevic}, {Jim{\'e}nez-Bail{\'o}n}, {La Franca}, {Smith},
  \& {Tosti}}]{Fup_Pen19}
{Pe{\~n}a-Herazo}, H.~A., {Massaro}, F., {Chavushyan}, V., {et~al.} 2019,
  \apss, 364, 85, \dodoi{10.1007/s10509-019-3574-4}

\bibitem[{{Salvetti} {et~al.}(2017){Salvetti}, {Mignani}, {De Luca}, {Marelli},
  {Pallanca}, {Breeveld}, {H{\"u}semann}, {Belfiore}, {Becker}, \&
  {Greiner}}]{J0744_Salvetti17}
{Salvetti}, D., {Mignani}, R.~P., {De Luca}, A., {et~al.} 2017, \mnras, 470,
  466, \dodoi{10.1093/mnras/stx1247}

\bibitem[{{Strader} {et~al.}(2014){Strader}, {Chomiuk}, {Sonbas}, {Sokolovsky},
  {Sand}, {Moskvitin}, \& {Cheung}}]{J0523_Strader14}
{Strader}, J., {Chomiuk}, L., {Sonbas}, E., {et~al.} 2014, \apjl, 788, L27,
  \dodoi{10.1088/2041-8205/788/2/L27}

\bibitem[{{Swihart} {et~al.}(2020){Swihart}, {Strader}, {Urquhart}, {Orosz},
  {Shishkovsky}, {Chomiuk}, {Salinas}, {Aydi}, {Dage}, \&
  {Kawash}}]{J2333_Swihart20}
{Swihart}, S.~J., {Strader}, J., {Urquhart}, R., {et~al.} 2020, \apj, 892, 21,
  \dodoi{10.3847/1538-4357/ab77ba}

\bibitem[{{Wang} {et~al.}(2020){Wang}, {Xing}, {Zhang}, {Boutsia}, {Wang}, {V},
  {Burdge}, {Coughlin}, {Duev}, {Kulkarni}, {Riddle}, \&
  {Serabyn}}]{Assoc20_Wang}
{Wang}, Z., {Xing}, Y., {Zhang}, J., {et~al.} 2020, \mnras, 493, 4845,
  \dodoi{10.1093/mnras/staa655}

\end{thebibliography}
 
\end{document}